\documentclass[seceq,preprint]{ptptex}
\usepackage{graphicx}
\usepackage{axodraw}

\renewcommand{\thefootnote}{\roman{footnote}}

\newcommand{\hs}[1]{\hspace*{#1 mm}}

\makeatletter
\def\tbcaption{\def\@captype{table}\caption}
\def\figcaption{\def\@captype{figure}\caption}
\makeatother

\newcommand{\feyngaugino}{
\begin{figure}[t]
\begin{center}
\hs{2}
\begin{picture}(500,80)(20,25)
\SetWidth{1.0}
\Photon(80,87)(80,97){2.5}{1}
\Photon(30,27)(50,37){2.5}{2}
\Photon(110,37)(130,27){2.5}{2}
\Line(50,37)(110,37)
\Line(50,37)(80,87)
\Line(110,37)(80,87)
\Text(63,107)[lb]{\large{$R^*,K$}}
\Text(135,12)[lb]{\large{$V$}}
\Text(15,12)[lb]{\large{$V$}}
\Text(105,62)[lb]{\large{$Q,Q'$}}
\Text(72,6)[lb]{\large{$\mbox{(a)}$}}
\Photon(250,87)(250,97){2.5}{1}
\Photon(280,37)(300,27){2.5}{2}
\Photon(220,37)(200,27){2.5}{2}
\ArrowLine(220,37)(280,37)
\ArrowLine(250,87)(220,37)
\ArrowLine(280,37)(250,87)
\Text(240,107)[lb]{\large{$b_m,a_m$}}
\Text(305,12)[lb]{\large{$v_n$}}
\Text(185,12)[lb]{\large{$v_l$}}
\Text(275,62)[lb]{\large{$\chi,\chi'$}}
\Text(242,6)[lb]{\large{$\mbox{(b)}$}}
\Line(420,87)(420,97)
\ArrowLine(450,37)(470,27)
\ArrowLine(370,27)(390,37)
\Line(420,87)(390,37)
\Line(390,37)(450,37)
\Line(450,37)(420,87)
\Text(400,107)[lb]{\large{$M^*,K_i F^i$}}
\Text(475,12)[lb]{\large{$\lambda$}}
\Text(355,12)[lb]{\large{$\lambda$}}
\Text(445,62)[lb]{\large{$A'$}}
\Text(414,6)[lb]{\large{$\mbox{(c)}$}}
\end{picture}
\end{center}
\hs{5}
\caption{
{\bf(a)} Anomalous superdiagram related to super-Weyl symmetry.
{\bf(b)} The well-known anomaly triangle diagram which is the component diagram
 of (a).
{\bf(c)} Feynman diagram deriving the gaugino mass which is also the
 component of (a). In this diagram only the massive fields as PV ones encircle. 
}
\label{feyngaugino}
\end{figure}
} 
 
\newcommand{\feynscalar}{
\begin{figure}[t]
\begin{center}
\begin{picture}(210,80)(0,15)
\SetWidth{1.0}
\Line(100,55)(80,75)
\Line(120,55)(140,75)
\CArc(110,35)(22.36,117,477)
\Photon(90,25)(70,5){2.5}{2}
\Photon(130,25)(150,5){2.5}{2}
\Line(50,5)(170,5)
\Text(60,85)[lb]{\large{$R^*,K$}}
\Text(130,85)[lb]{\large{$R,K$}}
\Text(175,0)[lb]{\large{$Q$}}
\Text(25,0)[lb]{\large{$Q^\dag$}}
\Text(145,15)[lb]{\large{$V$}}
\Text(105,15)[lb]{\large{$Q'$}}
\end{picture}
\end{center}
\hs{5}
\caption{
Feynman diagram deriving the scalar mass if the external
matter fields are the scalars $A_i$.
}
\label{feynscalar}
\end{figure}
} 

\newcommand{\figE}{
\begin{figure}[t]
\includegraphics*[scale=1.00]{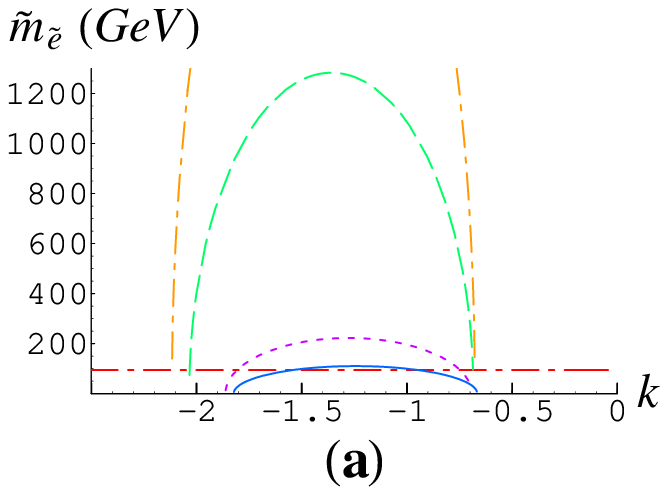}
\includegraphics*[scale=1.00]{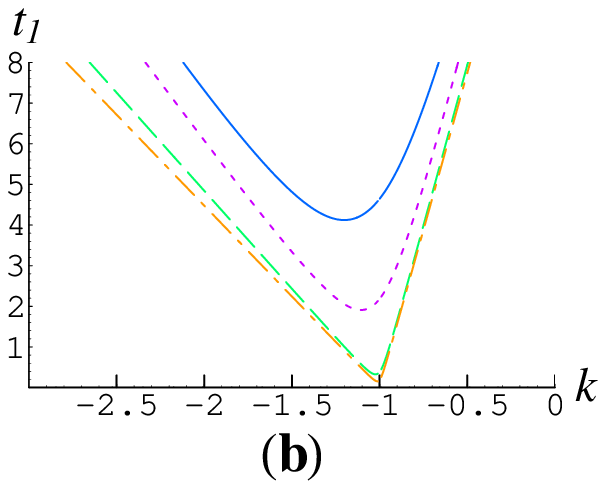}
\caption{
Fig.(a) shows the right handed slepton mass as the function of $k$ at $t_1=5$.
Fig.(b) shows the contour of $k$ and $t_1$ giving the mass the value of 94GeV, which is the present experimental lower bound,
where the allowed region is above the each line.
In both Figs, solid, dotted, broken and dashed lines correspond to 50, 100, 500 and 1000TeV, respectively.
}
\label{figE}
\end{figure}
}

\newcommand{\figh}{
\begin{figure}[t]
\includegraphics*[scale=1.00]{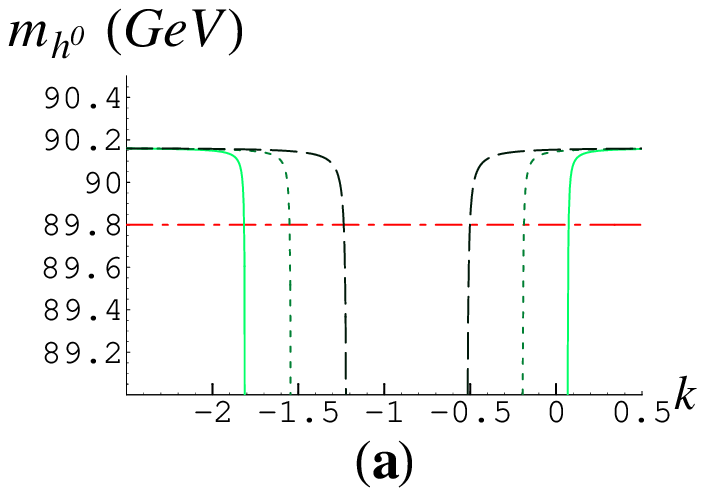}
\includegraphics*[scale=1.00]{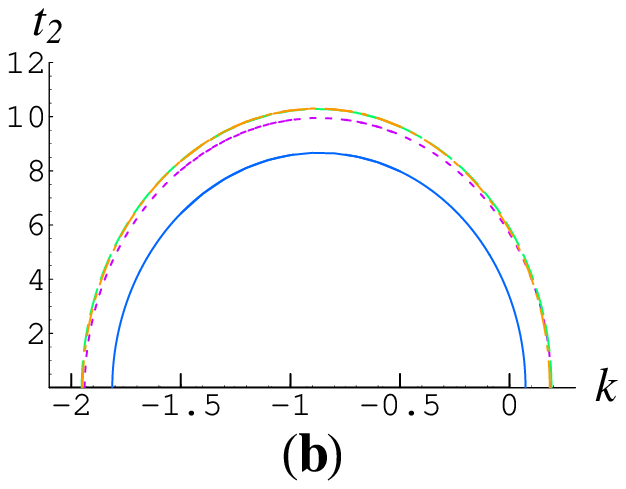}
\caption{
\textbf{(a)} The neutral scalar Higgs mass $\tilde{m}_{h^0}$ as the function of $k$ at $t_2=0$, 6 and 8, fixing $\frac{M}{3}=50$TeV.
\textbf{(b)} The contour of $t_2$ and $k$ giving $\tilde{m}_{h^0}$ the experimental lower bound 89.8GeV, where $\frac{M}{3}$ is varied as 50, 100, 500 and 1000TeV, and the allowed region is over the line as the same Fig.\ref{figE}(b).
}
\label{figHA}
\end{figure}
}

\newcommand{\figMu}{
\begin{figure}[t]
\includegraphics*[scale=1.10]{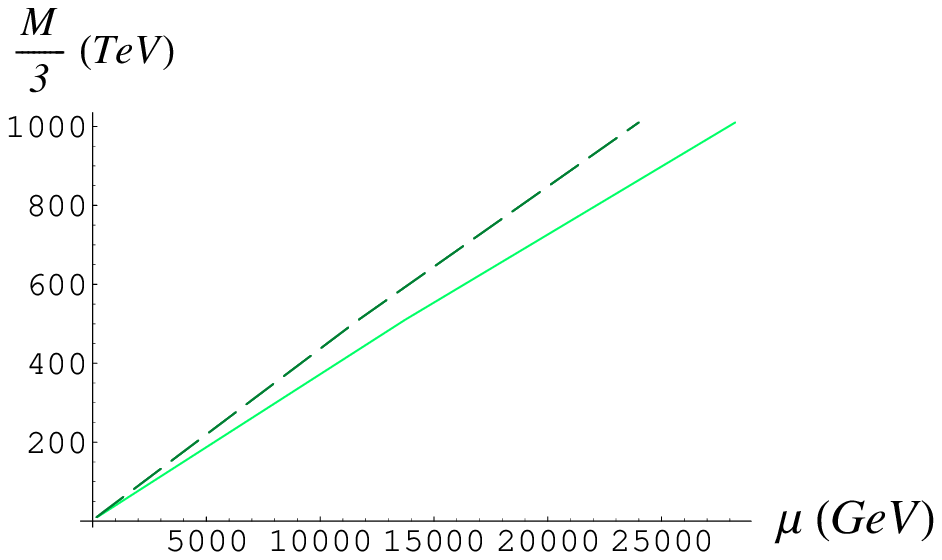}
\caption{
The behavior of $\frac{M}{3}$ as a function of $\mu$ by putting $v =245$GeV in (\ref{Vrelation}).
The solid and broken lines correspond to $(k,t_1,t_2)=(-1.6,3,8)$, and $(k,t_1,t_2)=(-1.8,4,5)$, respectively.
}
\label{figMu}
\end{figure}
}

\newcommand{\figt}{
\begin{figure}[t]
\includegraphics*[scale=1.00]{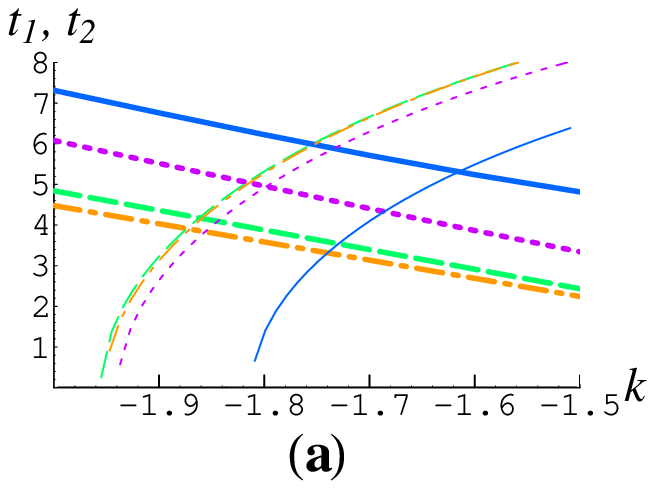}
\includegraphics*[scale=1.00]{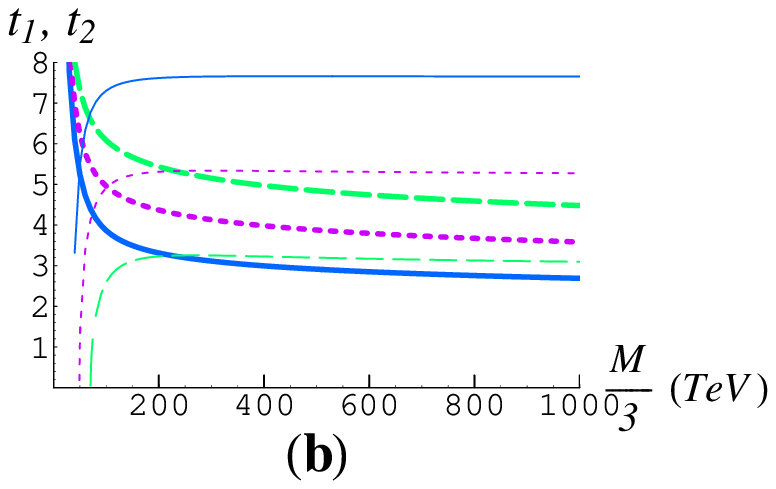}
\caption{
\textbf{(a)} The enlarged graph of Figs.\ref{figE}(b) and \ref{figHA}(b) at the regions $-2<k<-1.5$, where $t_1$ is drawn by bold lines
\textbf{(b)} The graph as the function of $\frac{M}{3}$ instead of $k$, fixing $k$ equal to $-1.6$, $-1.8$ and $-1.9$, which correspond to solid, dotted and broken lines.
}
\label{figt}
\end{figure}
}

\newcommand{\figL}{
\begin{figure}[t]
\includegraphics*[scale=1.00]{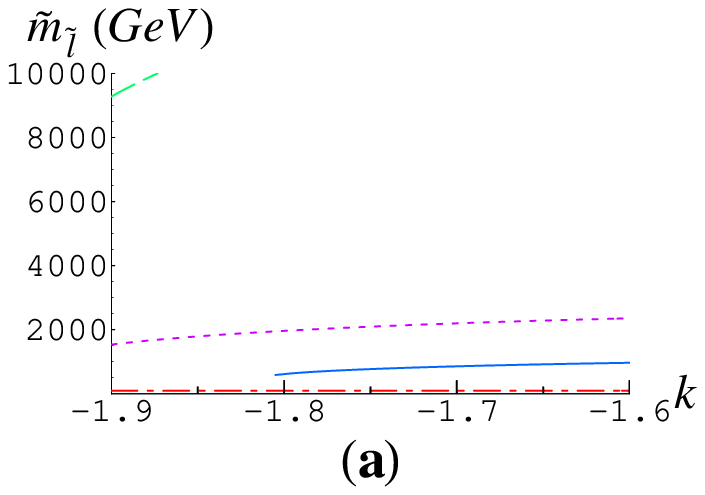}
\includegraphics*[scale=1.00]{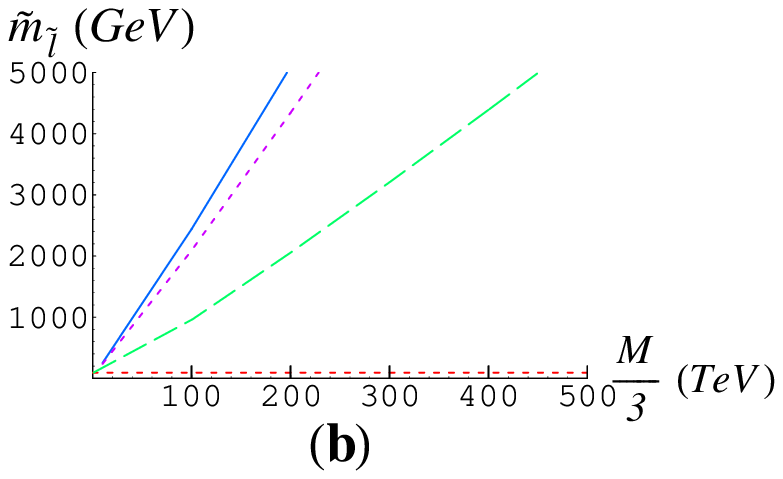}
\caption{
Left handed slepton $\tilde{m}_{\tilde{l}}$ masses.
In \textbf{(a)} as the function of $k$, solid, dotted and broken lines correspond to $\frac{M}{3}=50$, 100 and 500TeV, respectively.
In \textbf{(b)} as the function of $\frac{M}{3}$, these lines correspond to $k=-1.6$, $-1.8$ and $-1.9$.
The dashed line denote the experimental lower bound.
}
\label{figL}
\end{figure}
}

\newcommand{\figH}{
\begin{figure}[t]
\includegraphics*[scale=1.00]{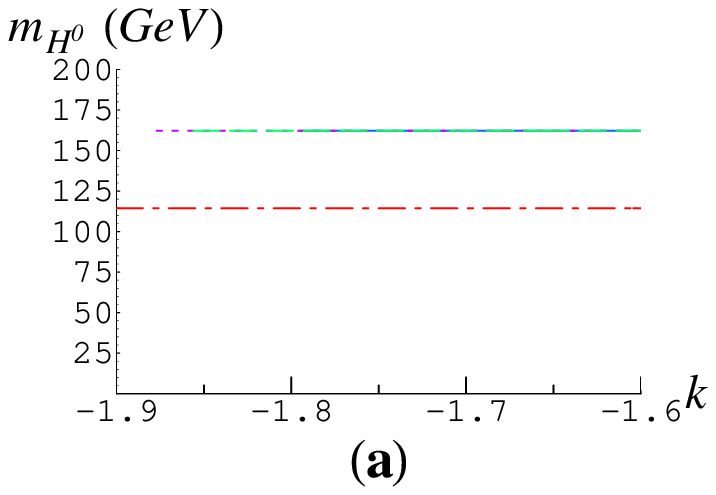}
\includegraphics*[scale=1.00]{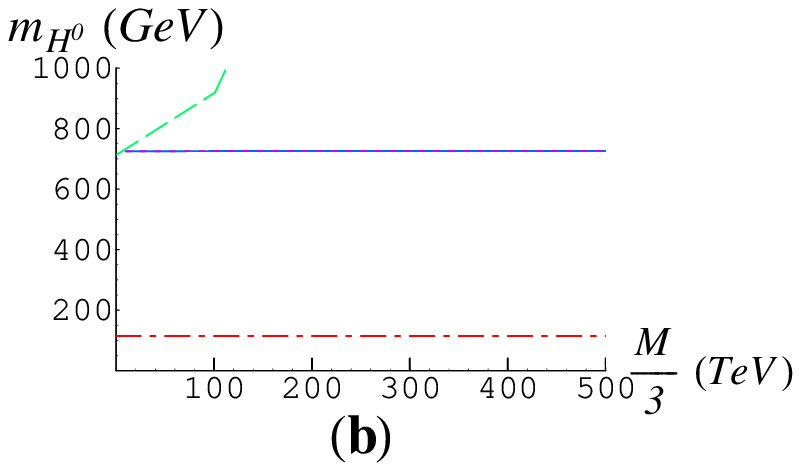}
\caption{
We illustrate neutral scalar Higgs $m_{H^0}$ masses.
In \textbf{(a)} the three lines degenerate.
In \textbf{(b)} the two lines, corresponding to $k=-1.6$ and $-1.8$, also degenerate.
}
\label{figH}
\end{figure}
}

\newcommand{\figChia}{
\begin{figure}[t]
\includegraphics*[scale=1.00]{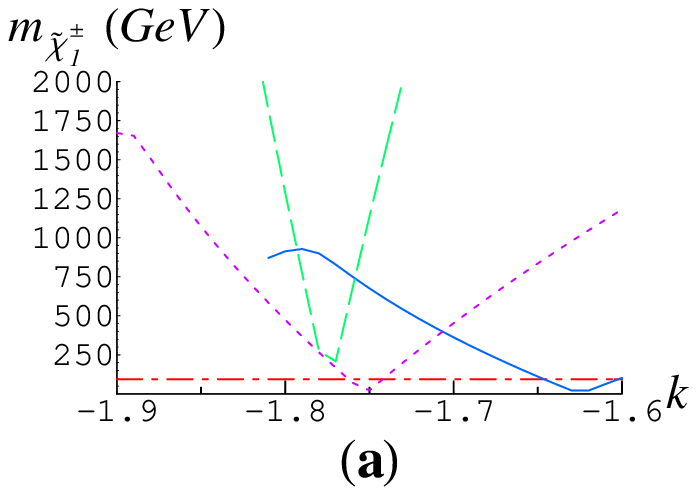}
\includegraphics*[scale=1.00]{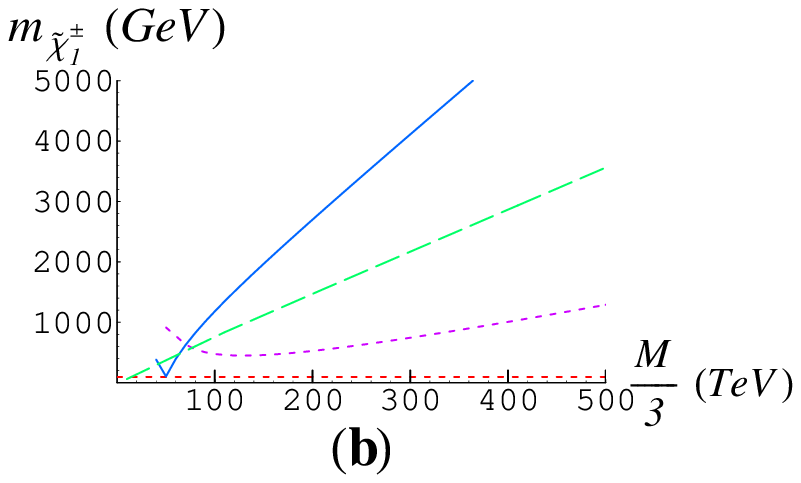}
\caption{
Chargino $m_{\tilde{\chi}^\pm_1}$ masses.
In \textbf{(a)} as the function of $k$, solid, dotted and broken lines correspond to $\frac{M}{3}=50$, 100 and 500TeV, respectively.
In \textbf{(b)} as the function of $\frac{M}{3}$, these lines correspond to $k=-1.6$, $-1.8$ and $-1.9$.
The dashed line denote the experimental lower bound.
}
\label{figChia}
\end{figure}
}

\newcommand{\figChib}{
\begin{figure}[t]
\includegraphics*[scale=1.00]{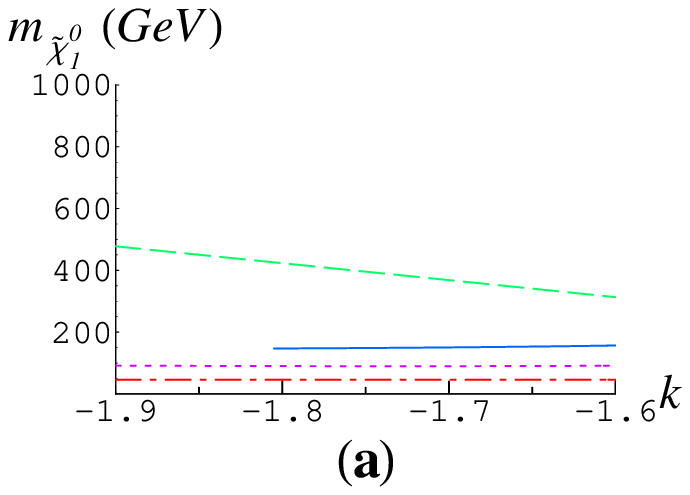}
\includegraphics*[scale=1.00]{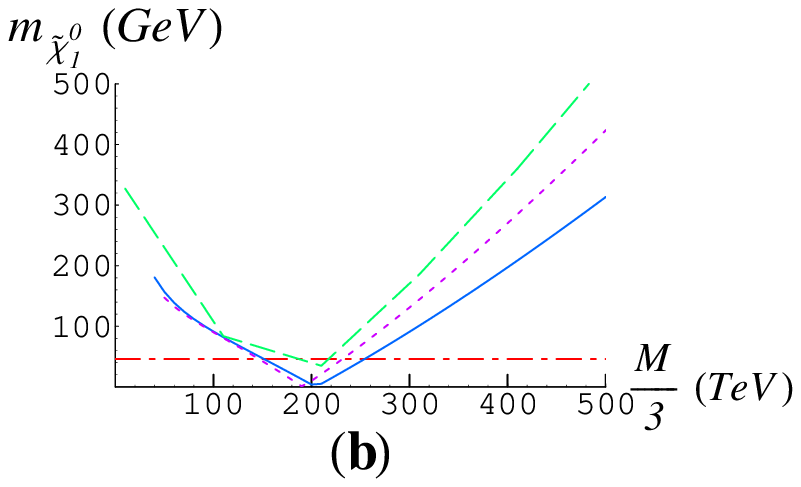}
\caption{
Neutralino $m_{\tilde{\chi}^0_1}$ masses.
}
\label{figChib}
\end{figure}
}

\newcommand{\MassBound}{
\begin{table}[t]
\caption{Experimental mass bound.}\label{MassBound}
\begin{tabular}{lc|c}
MSSM particles   &   & Mass Bound (GeV) \\
\hline\hline
Selectron      & $m_{\tilde{E}}$                            & $>$73 \\
Smuon      & $m_{\tilde{\mu}}$                            & $>$94 \\
Stau      & $m_{\tilde{\tau}}$                            &$>$81.9 \\
Charged Higgs              & $m_{H^\pm}$               &$>$79.3   \\
Pseudo Higgs                & $m_{A^0}$                     & $>$90.4   \\
Neutral Higgs                &  $m_{h^0},m_{H^0}$                    &$>$89.8,   $>$114.4\\
Chargino                        & $m_{\tilde{\chi}^\pm_1},m_{\tilde{\chi}^\pm_2}$         & $>$94, $\sim$\\
Neutralino                     & $m_{\tilde{\chi}^0_1},m_{\tilde{\chi}^0_2},m_{\tilde{\chi}^0_3},m_{\tilde{\chi}^0_4}$           &$>$46, $>$62.4, $>$99.9, $\sim $ 
\end{tabular}
\end{table}
}

\notypesetlogo                       
\preprintnumber[3cm]{OU-HET 508\\
hep-ph/0412233\\
March 4, 2005}

\title{
K{\"a}hler~Anomaly~Effect in~Anomaly~Mediation~Framework
}

\author{
Koske \textsc{Nishihara}\footnote{e-mail:koske@het.phys.sci.osaka-u.ac.jp}%
}

\inst{
Graduate~School~of~Science,~Osaka~University, 
Toyonaka,~Osaka~560-0043,~Japan}

\abst{
We extend the anomaly mediation mechanism by including the effect due to K{\"a}hler anomaly.
We give a general method analyzing the soft breaking terms in MSSM by introducing a set of parameters.
One of the parameters describes the magnitude of the K{\"a}hler contribution as opposed to the ordinary super-Weyl contribution.
The other parameters come from the so-called bilinear terms which are added to the K{\"a}hler potential in order for the gauge singlet scalar masses squared to be positive.
We explore allowed regions of these parameters by considering present experimental bound and present a new way of looking at model building.
}

\begin{document}

\maketitle

\setcounter{footnote}{0}
\renewcommand{\thefootnote}{\arabic{footnote}}

\section{Introduction}
As the theory beyond the Standard Model it has been believed that the Minimal Supersymmetric Standard Model (MSSM) is one of the most promising candidate.
Since supersymmetry (SUSY) is broken apparently in the real world, various scenarios for the SUSY breaking mechanism have been studied so far, such as gravity mediation \cite{Nilles:1983ge}, gauge mediation \cite{Giudice:1998bp} and so forth.
Several years ago, a novel type of SUSY breaking mechanism was proposed based on the super-Weyl anomaly, which is referred to as anomaly mediation mechanism \cite{Giudice:1998xp,Randall:1998uk}.
The most appealing aspect of anomaly mediation is its unique predictability of soft breaking terms (SBT's) which are all determined simply by vacuum expectation value (VEV) of the auxiliary field $M$ of the gravity multiplet. 

In spite of its high predictability, the anomaly mediation scenario is not satisfactory in all of its details. Namely, scalar partners of leptons are given  negative mass squared. Various ideas have been proposed to solve this tachyonic slepton mass problem.
Those include:
(i)~new bulk field interactions in extra dimensions \cite{Randall:1998uk,Rattazzi:2003rj},
(ii)~threshold effect due to new gauge interactions \cite{Pomarol:1999ie}\tocite{Katz:1999uw}, 
(iii)~extra Yukawa couplings \cite{Allanach:2000gu,Chacko:1999am}, 
(iv)~Fayet-Iliopoulos $D$ terms of extra U(1) gauge interactions \cite{Jack:2000cd}\tocite{Luty:1999qc},
(v)~horizontal gauge symmetry \cite{Anoka:2003kn} and
(vi)~bilinear term effects \cite{Nelson:2002sa}.
Only after solving the tachyonic slepton problem, we can confront the SUSY breaking mechanism with experimental data.

In Ref.\citen{Nishihara:2003bx}, we have proposed a new way of solving the slepton problem.
Namely we have extended the anomaly mediation taking account of not only super-Weyl anomaly but also K{\" a}hler one.
The original proposal of the anomaly mediation is a restricted case in which the K{\" a}hler anomaly effect vanishes.
There is, however, no compelling reason that we can neglect potentially important K{\" a}hler contribution.
The analysis in Ref.\citen{Nishihara:2003bx}, in fact, shows that the K{\" a}hler anomaly contribution can give positive mass squared to gauge non-singlet sleptons.
It has turned out, however, that the K{\"a}hler anomaly contribution is not effective for gauge singlet slepton, and further improvement has been proposed in Ref.\citen{Nishihara:2003bx} by including bilinear terms in K{\" a}hler potential \cite{Nelson:2002sa}.
The purpose of the present paper is to give a more detailed analysis of the idea of Ref.\citen{Nishihara:2003bx}.
We introduce a set of parameters to keep our analysis as comprehensive as possible.
One of the parameters describes the magnitude of the K{\"a}hler anomaly contribution compared with the ordinary super-Weyl anomaly and the other parameters represent the effect of the bilinear terms.

The present paper is organized as follows.
In section \ref{SecFormulae}, after a brief review of the anomaly mediation mechanism, we derive new mass formulae which include the effects due to K{\"a}hler anomaly and bilinear terms.
In the section the parameters describing these new effects are introduced.
In section \ref{MassSpectrum}, we analyze the allowed regions of the new parameters using our formula in the case of MSSM gauge group.
We will look for the possibility to build various models.
Section \ref{SecConclusion} is devoted to conclusions.

\section{Anomaly Mediation with K{\"a}hler Anomaly Effect}\label{SecFormulae}
\subsection{Anomaly Mediation}
In conventional studies of MSSM, the SBT's are given just by hand,
while the anomaly mediation mechanism predicts gaugino masses, scalar masses and trilinear coupling as follows,
\begin{align}
M_{\lambda_l} & = \frac{\beta_{g_l} }{g_l}  \frac{M}{3} ^* \label{AMgauginomass},\\
m^2_i  &= -\frac{1}{4}   \left(\frac{\partial\gamma_i}{\partial g}\beta_g,
               +\frac{\partial\gamma_i}{\partial y}\beta_y  \right) \left| \frac{M}{3}\right|^2 \label{AMscalarmass},\\
A^{ijk}  & =  \frac{1}{2} (\gamma_i+\gamma_j+\gamma_k)y^{ijk} \frac{M}{3} ^* ,
\end{align}
where indices $i$, $j$ and $k$ label the chiral matter multiplets and $l$ labels the gauge group.
These SBT's are all determined in terms of the beta functions of the gauge coupling $\beta _{g_l}$ of the gauge group $G_l$, the beta functions of the Yukawa coupling $\beta _{y}$ and the anomalous dimensions $\gamma _{i}$ of the $i$th chiral multiplet field.
These are all parameterized by the VEV $M$, which is the auxiliary field of the gravity multiplet.
There is no other mass scale other than $M$, and the predictions are unique.

The so-called tachyonic slepton mass problem is that scalar partners of leptons are given negative mass squared. Expanding  (\ref{AMscalarmass}) to the lowest order of $g$ we have
\begin{eqnarray}
m^2_{i}  \approx    c b g^4  \left|\frac{M}{3}\right|^2 ,\label{eqnega}
\end{eqnarray}
where $c$ and $b$ are the one loop coefficient of $\gamma$ and $\beta$.
Since for all gauge representations $c$ is negative, the mass squared gets the negative value in the case of asymptotically non-free gauge, $b>0$, as the slepton.
In the next subsection we show extended formulae for (\ref{AMgauginomass}) and (\ref{AMscalarmass}) by taking account of K{\" a}hler anomaly,
thereby pointing out a possibility to solve the tachyonic slepton problem.

\subsection{Super-Weyl-K{\" a}hler Anomaly}
We derive SBT's diagrammatically extending the formulae (\ref{AMgauginomass}) and (\ref{AMscalarmass}). 
The minimal supergravity coupled with matter and gauge multiplets is invariant under the simultaneous super-Weyl and K{\" a}hler transformations, but not under each separately. 
On the quantum level this symmetry is broken down, so the theory is anomalous \cite{LopesCardoso:1993sq}. 
Not only Weyl part but also K{\" a}hler one of the anomaly should potentially contribute to SBT's.
This idea considering these additional contributions was put forward for the case of the gaugino term in Ref.\citen{Bagger:1999rd} and for several SBT's in Ref.\citen{Gaillard:2000fk}.
Gaillard and Nelson \cite{Gaillard:2000fk} in particular made use of the K{\" a}hler U(1) superspace formalism \cite{Binetruy:2000zx} to compute one-loop quantum effect. 
In the same manner as the original anomaly mediation case, we assume that the hidden sector dose not has direct coupling to MSSM fields and the effect through the anomaly to SBT's is dominant.

As usual the super-Weyl-K{\" a}hler anomaly arises through the anomalous triangle diagram Fig.\ref{feyngaugino}(a), in which $R$ is the chiral superfield containing the scalar curvature and $K$ is the K{\"a}hler potential.
Cardasso and Ovrut \cite{LopesCardoso:1993sq} calculated Fig.\ref{feyngaugino}(b), 
which is one of the component diagrams of Fig.\ref{feyngaugino}(a) and the connection of the super-Weyl and K{\"a}hler transformations are denoted respectively by $b_m$ and $a_m$, 
and they guessed the super-Weyl-K{\" a}hler anomaly formulae of the superfield form.
We must calculate Fig.\ref{feyngaugino}(c) which is the other component of (a) to get gaugino mass.  
Though we can guess the result of calculation from Fig.\ref{feyngaugino}(a), we want to calculate it directly to apply the method to the scalar mass.
In Fig.\ref{feyngaugino}(c) only massive fields encircle the diagram because massless fields don't couple to $M$ and $K_i F^i$, where $K_i= \frac{\displaystyle{\partial K}}{\displaystyle{\partial A_i}}$, so in the massless theory these terms don't contribute to the calculation.
A mass scale, however, has to be introduced to regularize the theory even if the theory has only massless fields.
In the following we use the Pauli-Villars (PV) regularization, and not only the matter fields $Q_i (=A_i+\sqrt{2}\theta \chi_i +\theta\theta F_i)$, but PV fields $Q'_i$ encircle the triangle Fig.\ref{feyngaugino}(a). 
Thus PV fields encircle in the diagram (c), which would produce gaugino mass $m_\lambda$ if the auxiliary field $M$ is given VEV.
 
\feyngaugino

Let us then calculate gaugino mass $m_\lambda$ coming from
Fig.\ref{feyngaugino}(c). The Lagrangian of supergravity coupled with
matter and gauge fields is\footnote{In the following argument we use the notation in Ref.\citen{Wess:cp}.}
\begin{equation}
\mathcal{L} =\mathcal{L}_M+\mathcal{L}_G, \label{wholeL}
\end{equation}
\begin{align}
\mathcal{L}_M = & \frac{1}{\kappa^2} \int \! d^2 \Theta\  2\mathcal{E}
\! \left[ \frac{3}{8}(\bar{\mathcal{D}}\bar{\mathcal{D}}-8R) \exp\left\{-\frac{\kappa^2}{3}K \right\} +\Lambda_i \bar{Q}'_i Q'_i \right] + h.c. \\
&K= Q^{\dagger}_i e^{2V} Q_i + Q'^{\dagger}_{i} e^{2V} Q'_{i}.\nonumber\\
\mathcal{L}_G = & \int \! d^2 \Theta\  2\mathcal{E}
\! \left[  \frac{1}{4g^2}  {W^a}^\alpha W^a_\alpha   - \frac{1}{8}(\bar{\mathcal{D}}\bar{\mathcal{D}}-8R)   \Phi'^{\dagger} e^{2V} \Phi'  +\Lambda_\Phi \bar{\Phi}' \Phi' \right] + h.c.
\end{align}
Here,  $Q'_i$'s  are
chiral PV fields  of the same representation as matter fields
and $\Phi' (= A'_\Phi+\sqrt{2}\theta \chi'_\Phi +\theta\theta F'_\Phi)$
are those of the adjoint representation. 
The PV fields are distinguished by prime " $'$ " from ordinary fields.  $\bar{Q}$ and 
$\bar{\Phi}$ are the conjugate representation chiral fields of $Q$ and $\Phi$
respectively. 
Furthermore, for the sake of simplicity we ignore Yukawa coupling. 

We would like to consider the low energy limit of (\ref{wholeL}). 
Taking the flat limit for component fields of gravity multiplet (graviton $e\rightarrow\delta$, gravitino $\psi\rightarrow 0$, auxiliary fields $b_n\rightarrow 0,M\rightarrow 0$ and gravitational coupling $\kappa^2 \rightarrow \infty$), eq.(\ref{wholeL}) becomes global SUSY Lagrangian $\mathcal{L}^{global} =\mathcal{L}_M^{global}+\mathcal{L}_G^{global}$, where 
\begin{align}
\mathcal{L}^{global}_M =& \int \! d^4 \theta\  
\! \left[   Q^{\dagger}_i e^{2V} Q_i + Q'^{\dagger}_{i} e^{2V} Q'_{i} \right] 
+\int \! d^2 \theta\  \Lambda_i \bar{Q}'_i Q'_i  + h.c. ,\\
 \mathcal{L}^{global}_G = &  \int \! d^2 \theta\  
\!   \frac{1}{4g^2}  {W^a}^\alpha W^a_\alpha   + h.c. \nonumber\\
&+ \int \! d^4 \theta\   \Phi'^{\dagger} e^{2V} \Phi'
+\int \! d^2 \theta\ \Lambda_\Phi \bar{\Phi}' \Phi'  + h.c.
\end{align}

However, the auxiliary field $M$ and K{\" a}hler potential $K$ may be given VEV if SUSY is broken down by the VEV of hidden sector fields. In this case we can take the limit,  $e\rightarrow\delta$, $\psi\rightarrow 0$, $b_n\rightarrow 0$, while giving non-zero VEV to $M$,
\begin{eqnarray}
\mathcal{L} \rightarrow  \mathcal{L}^{global} + \left[ -\Lambda_i \left\{\frac{M}{3}^* - \frac{2}{3} \kappa^2 K_i F^i \right\}\! \bar{A}'_i A'_i -\Lambda_\Phi \frac{M}{3}^* \! \bar{A}'_\Phi A'_\Phi + h.c. \right]. \label{LlowenargylimitWK}
\end{eqnarray}
Here the terms proportional to $M^*$ and $\kappa^2 K_i F^i$ appear in addition to the global SUSY Lagrangian $\mathcal{L}^{global}$. 
The point which we should stress is that the effect of $M^*$ and $\kappa^2 K_i F^i$ appears only in the PV fields part. 
As we mentioned above the contribution comes from only the massive PV field encircling the triangle diagram.
One should also note that $\kappa^2 K_i F^i$ is accompanied by only the matter PV fields $A'_i$, but not the gauge PV fields $A'_\Phi$.  This means the term $\kappa^2 K_i F^i$ contributes only to the loop diagram encircled by $A'_i$.

Firstly let us note that, were it not for the term $\kappa^2 K_i F^i$, Fig.\ref{feyngaugino}(c) would give us the gaugino mass
\begin{eqnarray}
M_{\lambda_l} = \frac{\alpha_l(\mu)}{4\pi} \left[ -3C_{G _l}+\sum_i T^l_{R_i}  \right] \frac{M}{3} ^*, \label{lowestgauginomass}
\end{eqnarray}
where $C_G$ is the value of the quadratic Casimir operator of the
adjoint representation, $T_R$ is the Dynkin index associated with the
matter representation and $\mu$ means the low energy scale we could observe. 
This is the lowest order term of pure anomaly mediation gaugino mass (\ref{AMgauginomass}).

In the case that the term $\kappa^2 K_i F^i$ 
is included,
the calculation is straightforward because of a simple replacement rule \cite{Gaillard:2000fk,Nishihara:2003bx}.
As is clear from (\ref{LlowenargylimitWK}), we replace $\frac{M}{3}^*$ coming from the matter PV field by
\begin{align}
\frac{M}{3}^* \rightarrow \frac{M}{3}^* - \frac{2}{3} \kappa^2 K_i F^i, \label{replacing}
\end{align}
while the gauge PV field contribution is left untouched.
Therefore we end up with
\begin{align}
M_{\lambda_l} = \frac{\alpha_l(\mu)}{4\pi} \left[  -3C_{G _l} \frac{M}{3}^*  + \sum_i T^l_{R_i} \left\{ \frac{M}{3}^* - \frac{2}{3} \kappa^2 K_j F^j \right\} \right]. \label{gauginomass}
\end{align}

\feynscalar

The masses $m_i$ of the scalar fields $A_i$ can be also derived similarly. In this case, however, the scalar masses arise on the two-loop level through the diagram of the type depicted in Fig.\ref{feynscalar}. In the component diagram of Fig.\ref{feynscalar}, whose external matter fields are the scalars $A_i$, the direct evaluation gives us the scalar masses,
\begin{eqnarray}
m^2_i =\sum_l \frac{\alpha_l^2(\mu)}{(4\pi)^2} C^l_{R_i}\left[3C_{G_l} -\sum_i T^l_{R_i} \right]\left| \frac{M}{3}\right|^2, \label{lowestscalarmass}
\end{eqnarray}
without the $\kappa^2 K_i F^i$ terms.
Summation is taken over all the gauge group indices. 
This result agrees in the lowest order with (\ref{AMscalarmass}), too.
Employing the same replacement rule (\ref{replacing}) we get the scalar mass formula with the K{\" a}hler anomaly effect,
\begin{eqnarray}
m^2_i = \sum_l \frac{\alpha_l^2(\mu)}{(4\pi)^2} 2C^l_{R_i}\left[3C_{G _l}  \left| \frac{M}{3}\right|^2 -\sum_j T^l_{R_j}  \left| \frac{M}{3}^* - \frac{2}{3} \kappa^2 K_k F^k \right|^2\right]. \label{scalarmass}
\end{eqnarray}

Although we don't go into the details of the hidden sector, it is natural to assume that $\frac{M}{3}^*$ and $-\frac{2}{3}\kappa^2 K_i F^i$ are on the same order of magnitude as is argued in Ref.\citen{Bagger:1999rd}. 
We are thus led to introduce a parameter $k$ of the order of unity via
\begin{eqnarray}
k= - \frac{\frac{2}{3}\kappa^2 K_i F^i}{\frac{1}{3}\displaystyle{M^*}}.\label{kdefinition}
\end{eqnarray}
In terms of this new parameter the formulae (\ref{gauginomass}) and (\ref{scalarmass}) are rewritten as
\begin{align}
M_{\lambda_l} = &\frac{\alpha_l(\mu)}{4\pi} \left[  -3C_{G _l}  + \sum_i T^l_{R_i} ( 1 +k ) \right] \frac{M}{3}^*\label{ReplacedGauginomass},\\
m^2_i = &\sum_l \frac{\alpha_l^2(\mu)}{(4\pi)^2} 2C^l_{R_i}\left[3C_{G _l}  -\sum_j T^l_{R_j} ( 1 +k )^2\right]\left| \frac{M}{3}\right|^2, \label{ReplacedScalarmass}
\end{align}

\subsection{Bilinear Terms}
Eq.(\ref{ReplacedScalarmass}) tells us that gauge non-singlet sleptons acquire positive mass square if $3C_{G _l}>\sum_i T^l_{R_i}(1+k)^2$.
For singlet slepton, however, the positivity of $m_i^2$ is not fulfilled because of $C_G=0$ and $T_R>0$, irrespectively of the value of $k$.
Slight modification of the model is necessary to remedy this defect.
We use the idea in Ref.\citen{Nelson:2002sa} of introducing extra fields $\phi_n$ $(n=1,2,3,\cdots)$ with bilinear terms into the K{\" a}hler potential,
\begin{equation}
K_{\phi}= \phi_n^\dagger \phi_n + \bar{\phi}_n^\dagger \bar{\phi}_n - c_n \bar{\phi}_n\phi_n- c_n^* \bar{\phi}_n^\dagger \phi_n^\dagger. \label{addtionalK}
\end{equation}
Here the coefficients $c_n$ are in general complex numbers and are free parameters.
For the sake of simplicity, however, we assume $c_n$ are all real and on the order of unity. 
Using above K{\" a}hler potential the following terms are added to Lagrangian due to the bilinear terms,
\begin{equation}
\Delta \mathcal{L}_\phi=\int\! d^2\theta \left[ c_{n} \frac{M^*}{3} \bar{\phi}_n\phi_n\right]  + c_{n} \left| \frac{M^*}{3}\right|^2  \bar{A}_{\phi_n} A_{\phi_n}  +\mbox{h.c.}\label{Lwithbilinear}
\end{equation}
This means that the fields $\phi_n$ become massive, the mass is $c_n \frac{M}{3}=m_{c}^n$, due to bilinear terms.
Applying the same limit as in (\ref{LlowenargylimitWK}), the Lagrangian of additional fields is
\begin{eqnarray}
\mathcal{L}_{\phi} \rightarrow \mathcal{L}^{global}_{\phi} + \left[ -m_{c}^n \left\{-\frac{M}{3}^* - \frac{2}{3} \kappa^2 K_i F^i \right\}\! \bar{A}_{\phi_n}A_{\phi_n}  + h.c. \right] .\label{Llowenargylimitthreshold}
\end{eqnarray}
Comparing (\ref{LlowenargylimitWK}) with (\ref{Llowenargylimitthreshold}), we immediately notice a remarkable difference in sign in front of  $\frac{M}{3}^*$.  
In the original anomaly mediation without the bilinear terms,
heavy fields with the ordinary mass term do not contribute to the SBT.
However, because of this difference of sign the heavy fields $\phi_n$ with the bilinear terms discriminate themselves from ordinary massive fields, thus the threshold effects survive the low energy limit.
We can just calculate the bilinear effect in the same way as before with Feynman diagram. 
The result is to add the following term
\begin{eqnarray}
\begin{split}
\Delta M_{\lambda_l} &= \frac{\alpha_l(\mu)}{4\pi} 2 \left[ \sum_n \left( T^l_{R_{\phi_n}}+T^l_{R_{\bar{\phi}_n} } \right)\right]  \frac{M}{3}^* \label{deltagaugino}\\
&= \frac{\alpha_l(\mu)}{4\pi}   2 t_l  \frac{M}{3}^*, 
\end{split}\label{gauginomassthreshold}
\end{eqnarray}
to (\ref{ReplacedGauginomass}), where 
\begin{eqnarray}
t_l = \sum_n (T^l_{R_{\phi_n}}+T^l_{R_{\bar{\phi}_n}} ).\label{tbilinear}
\end{eqnarray}

It is generally believed that the SBT's are insensitive to heavy fields.
Eq.(\ref{gauginomassthreshold}) shows, however , that the heavy fields $\phi_n$ affects the gaugino mass directly.
It is easy to explain this fact by using a compensator $\Phi_c$.
In this formalism the effect to SBT's is interpreted as a result of alteration under the conformal rescaling $\Phi_c Q_i \rightarrow Q_i$.
The effect of the effective gauge coupling from heavy fields $S_{h}$ is given by \cite{Giudice:1997ni}
\begin{eqnarray}
S_h \approx \frac{b_h}{16\pi^2}\ln \frac{m_h}{\Lambda},\label{effectivegaugeS}
\end{eqnarray}
where $m_h$ is the mass of the heavy field and $b_h$ is the first coefficient of the $\beta$ function due to the heavy field.
Because the ordinary mass transforms $m \rightarrow \Phi_c m$ under the conformal rescaling, eq.(\ref{effectivegaugeS}) is invariant due to 
\begin{eqnarray}
\ln\frac{m_h}{\Lambda} \rightarrow \ln\frac{\Phi_c m_h}{\Phi_c\Lambda}=\ln\frac{m_h}{\Lambda} \label{Ccancellation}
\end{eqnarray}
This is the explanation in the compensator formalism why heavy fields don't contribute to SBT's.
In the mass caused by the bilinear terms, however, the dependence on the compensator is different.
The transformation of the mass is given by $m_c \rightarrow \Phi_c^{-1} m_c$ as a result of difference of the Weyl weight from the ordinary mass.
This difference appears in our formalism as the difference of sign in eq.(\ref{Llowenargylimitthreshold}).
Thus instead of the cancellation (\ref{Ccancellation}) the variations are additive, i.e.,
\begin{eqnarray}
S_h= \frac{b_h}{16\pi^2}\ln\frac{m_c}{\Lambda} \rightarrow \frac{b_h}{16\pi^2}\ln\frac{\Phi_c^{-1} m_c}{\Phi_c\Lambda}
     =S_h - 2 \frac{b_h}{16\pi^2}  \ln \Phi.\label{Cnoncancellation}
\end{eqnarray}
The second term of the rightmost side of (\ref{Cnoncancellation}) induces the additional gaugino term (\ref{deltagaugino}).

Although we have added bilinear terms, the scalar masses don't receive contributions directly contrary to  (\ref{gauginomassthreshold}).
Explaining it in the compensator formalism, we observe that the wave-function renormalization $Z_Q$ is given by and transforms as \cite{Giudice:1997ni}
\begin{align}
Z_Q(\mu) =\ \ &Z_Q(\Lambda)\left[\frac{\alpha(\Lambda)}{\alpha(m_c)} \right]^{\frac{2C_R}{b'}}
                                             \left[\frac{\alpha(m_c)}{\alpha(\mu)} \right]^{\frac{2C_R}{b}} \\
     \longrightarrow \ &
					Z_Q(\Lambda)\left[\frac{\alpha(\Phi_c\Lambda)}{\alpha(\Phi_c^{-1}m_c)} \right]^{\frac{2C_R}{b'}}
                                             \left[\frac{\alpha(\Phi_c^{-1}m_c)}{\alpha(\mu)} \right]^{\frac{2C_R}{b}}\\
      =\ \ & Z_Q(\mu)\left[1+ \frac{\alpha(\Lambda)}{4\pi}b'\ln\Phi_c^{-4} \right]^{\frac{2C_R}{b'}}
        \left[
				\frac{1+ \displaystyle{\frac{\alpha(\Lambda)}{4\pi}}b'\ln\Phi_c^{-4}+\displaystyle{\frac{\alpha(\Lambda)}{4\pi}}b \ln\Phi_c^{2} }
					     {1+ \displaystyle{\frac{\alpha(\Lambda)}{4\pi}}b'\ln\Phi_c^{-4} }
													       \right]^{\frac{2C_R}{b}}\\								 
       =\ \ & Z_Q(\mu)\left[1 -4C_R\frac{\alpha(\Lambda)}{4\pi}  \ln\Phi_c \right]\\
	 =\ \ &
	  Z_Q(\Lambda)\left[\frac{\alpha(\Phi_c\Lambda)}{\alpha(\mu)} \right]^{\frac{2C_R}{b}},\label{Zcancellation}
\end{align}
under the rescaling, where $b'=b+b_h$.
The last line is the same as the case without the bilinear terms.
This indicates that there is no effect of the bilinear terms on the scalar masses at the energy scale $m_c$.
Because of the bilinear terms, however, SBT's are no longer on the anomaly mediation trajectory. 
Through the renormalization group (RG) flow the gaugino mass effect (\ref{gauginomassthreshold}) is transmitted to the scalar mass at the mass scale lower than $m_c$.
Thus we must solve the RG equation of the scalar mass squared and coupling constant
\begin{eqnarray}
\frac{d}{d \log\mu} m^2_i(\mu) &= & -8 \sum_l \frac{\alpha_l(\mu)}{4\pi} C^l_{R_i} \left| M_{\lambda_l}(\mu) \right|^2,\label{RGscalar}\\
\frac{d}{d \log\mu} M_{\lambda_l}(\mu) &= & \frac{2b_l}{4\pi} \alpha_l(\mu)M_{\lambda_l},\\
 \frac{d}{d \log\mu} \alpha_l (\mu)&= &\frac{2b_l}{4\pi} \alpha_l^2(\mu),
\end{eqnarray}
where $b_l$ is the one loop coefficient of $\beta$ function.
The initial conditions at $m_c$ are
\begin{align}
M_{\lambda_l}(m_c)  =&  \frac{\alpha_l(m_c)}{4\pi} \left[  -3C_{G_l}  + \sum_i T^l_{R_i} ( 1 +k )  +2 t_l \right] \frac{M}{3}^* ,\label{GauinoAtMc}\\ 
m^2_i(m_c)  =& \sum_l  \frac{\alpha_l^2(m_c)}{(4\pi)^2} 2C^l_{R_i}\left[3C_{G _l}  -\sum_j T^l_{R_j} ( 1 +k )^2\right]\left| \frac{M}{3}\right|^2.\label{ScalarAtMc}
\end{align}
After solving the RG equations, we end up with the formulae
\begin{align}
M_{\lambda_l}  =& \frac{\alpha_l(\mu)}{4\pi} \left[  -3C_{G_l}  + \sum_i T^l_{R_i} ( 1 +k )  +2 t_l \right] \frac{M}{3}^* , \label{GauginomassFormula}\\
m_i^2 =& \sum_l \left[ \left\{3C_{G_l}  -\sum_j T^l_{R_j} ( 1 +k )^2\right\} 2C^l_{R_i} \frac{{\alpha_l}^2(m_c)}{(4\pi)^2}  \right.\nonumber \\
 & \left. +\left\{ -3C_{G_l}  + \sum_j T^l_{R_j} ( 1 +k )  -2 t_l \right\}^2 2 \frac{C^l_R}{b_l}  \frac{\alpha_l^2(m_c) -\alpha_l^2(\mu)}{(4\pi)^2}\right] \left|\frac{M}{3}\right|^2 . \label{ScalarmassFormula}
\end{align}
In the case of $k$ = 0 this agrees with the result of Ref.\citen{Nelson:2002sa}

\section{Mass Spectrum}\label{MassSpectrum}
\subsection{MSSM and Experimental Bounds}

Here we discuss the mass spectra on the basis of (\ref{GauginomassFormula}) and  (\ref{ScalarmassFormula}).
We consider MSSM and the gauge group is $\mbox{SU}(3)_C \times \mbox{SU}(2)_L \times \mbox{U}(1)_Y$.
We denote the summations of the Dynkin index  (\ref{tbilinear}) for $\mbox{SU}(3)_C$, $\mbox{SU}(2)_L$ and $\mbox{U}(1)_Y$ by $t_3$,$t_2$ and $t_1$, respectively.
Since we are interested in implications to the $\mbox{SU}(2)_L \times \mbox{U}(1)_Y$ physics  such as those of the slepton and Higgs masses and the electroweak symmetry breaking, which are independent of $t_3$,  we consider four parameters, $M$, $k$, $t_1$ and $t_2$. 

The explicit values of brackets in the gaugino mass (\ref{GauginomassFormula}) are 
\begin{eqnarray}
-3C_{G _l}  + \sum_i T^l_{R_i} ( 1 +k )  -2 t_l = 
\left\{
\begin{split}     
 11(1 +k)  -2 t_1      &   \mbox{, for U}(1)_Y,\\
  -6+7 (1 +k)  -2 t_2  &   \mbox{, for SU}(2)_L.
\end{split}
\right.
\end{eqnarray}
Each term in the summation (\ref{ReplacedScalarmass}) of the scalar masses is
\begin{eqnarray}
2C_{R_i}\left[3C_{G _l}  -\sum_j T^l_{R_j} ( 1 +k )^2\right] =
\left\{
\begin{split}
 9-16(1 +k)^2 &\mbox{, for SU(2) doublet},\\
 -22(1 +k)^2 &\mbox{, for  SU(2) singlet}.
\end{split}
\right.
\end{eqnarray}

Up to this point now we have neglected the Yukawa couplings.
Numerically the top Yukawa coupling isn't allowed to be neglected. 
In this analysis only $m_{H_2}$ receives such an effect because of coupling to the top Yukawa.
To get an insight into the tachyonic slepton mass problem, however, it is more illuminating to proceed with approximation then to do precise computations numerically. 
For this reason, we replace the RG equation (\ref{RGscalar}) and the initial condition (\ref{ScalarAtMc}) by
\begin{eqnarray}
\frac{d}{d \log\mu} m^2_{H_2}(\mu) \approx 
   -6\frac{\alpha_2(\mu)}{4\pi} \left| M_{\lambda_2}(\mu) \right|^2
    -2\frac{\alpha_1(\mu)}{4\pi} \left| M_{\lambda_1}(\mu) \right|^2 
    + 6A_tA_t^\dag ,\label{RGmH2}
\end{eqnarray}
\begin{align}
m^2_{H_2}(m_c)  = &\left[
    \left\{ 9-\frac{21}{2}(1+k) \right\} \frac{\alpha^2_2(m_c)}{(4\pi)^2}
    -\frac{11}{2}(1+k)  \frac{\alpha^2_1(m_c)}{(4\pi)^2}\right.\nonumber\\
    &\ \ \ \ \ \ \ \ \ \ \ \ \ \left.+6 {y_t}^2(m_c)\left\{-\frac{16}{3}\frac{\alpha^2_3(m_c)}{(4\pi)^2}+ 6 {y_t}^2(m_c)\right\}\right] \left|\frac{M}{3}\right|^2 ,
\label{mH2init}
\end{align}
where, as we did in section \ref{SecFormulae} we assume $m_c$ is on the same order of magnitude as $\frac{M}{3}$, and $A_t$ is the $A$ term proportional to $y_t$,
\begin{eqnarray}
A_t(\mu) = 
    \left\{ -\frac{16}{3}\frac{\alpha_2(\mu)}{4\pi}+6\frac{y_t^2(\mu)}{(4\pi)^2} \right\} y_t(\mu) \frac{M}{3}^*.
\end{eqnarray}
In this equation stop mass or other scalar masses might contribute to the Higgs mass.
In the same way $A$ term running effect from gaugino masses might have contribution.
Considering the fact that these scalar masses are much smaller than gaugino masses in the initial condition, and we may assume that $A$ term running effect is not large and that we ignore these running effects in this rough estimation.
We estimate numerically with running from the scale of $\frac{M}{3}^*$ to the weak scale $\mu \approx 100$GeV.

\MassBound
Moreover, after obtaining the masses at the scale $\mu$ we must further diagonalize the mass matrix, taking linear combinations of fields of the same quantum number. 
The sets of fields that should mix up are: (i) left and right sleptons $(\tilde{e}_R,\tilde{l}_L)$, (ii) charged Wino and Higgsino $(\tilde{W}^\pm,\tilde{H}^\pm)$ which are called charginos $\tilde{\chi}_{1,2}^\pm$, (iii) Bino, neutral Wino and Higgsino $(\tilde{B},\tilde{W}^0,\tilde{H}^0_1,\tilde{H}^0_2)$ which are called neutralinos $\tilde{\chi}_{1,2,3,4}^0$ and (iv) charged and neutral Higgses $(H^\pm_1,H^\pm_2)$ and $(H^0_1,H^0_2)$, and refer to Ref.\citen{Haber:1984rc} for further details.
In the course of diagonalizing the mass matrix the MSSM parameters $\mu$ and tan$\beta$ come into the mass formulae.
Therefore we have six parameters $\mu$, tan$\beta$, $M$, $k$, $t_1$ and $t_2$.

Our analysis of the mass spectra must be given due consideration of the experimental bounds on the masses which are shown in Table \ref{MassBound}. 
Moreover in order to trigger the electroweak symmetry breaking, Higgs masses $m_{H1}$ and $m_{H2}$ must satisfy the following conditions,
\begin{eqnarray}
\begin{split}
\frac{1}{2}(\bar{m}^2_{H1} + \bar{m}^2_{H2}) > &  | B\mu | >  \sqrt{\bar{m}^2_{H1} \bar{m}^2_{H2}} & \mbox{for }& \bar{m}^2_{H1},\bar{m}^2_{H2}>0,\\
\frac{1}{2}(\bar{m}^2_{H1} + \bar{m}^2_{H2}) > &  | B\mu |  & \mbox{for }& \bar{m}^2_{H1}>0 ,\bar{m}^2_{H2}<0.
\end{split}
\label{ewbreakcondition}
\end{eqnarray}
The v.e.v of Higgs potential $v$ and Higgs mixing term $B\mu$ are expressed in terms of Higgs masses $m_{H1}$ and $m_{H2}$, $\mu$ and $\beta$,
\begin{align}
v^2 =& -\frac{4}{g_1^2+g_2^2} \left( \frac{\bar{m}^2_{H1} - \bar{m}^2_{H2}}{\cos 2\beta}+\bar{m}^2_{H1} + \bar{m}^2_{H2} \right),\label{Vrelation}\\
B\mu   =& - (\bar{m}^2_{H1} + \bar{m}^2_{H2} ) \sin 2\beta.
\end{align}
We assign the experimental value 245GeV to $v$, so the number of parameters is reduced to five.

\subsection{Parameters}

\figE
Our criterion of selecting the parameters is that we should minimize the number of additional fields $\phi_n$ with bilinear terms.
In other words $t_1$ and $t_2$ should be kept small, while the predicted masses satisfy the electroweak symmetry breaking conditions (\ref{ewbreakcondition}) and (\ref{Vrelation}) with the experimental bound.
For the sake of simplicity we fix tan$\beta = 15$.
Our remaining five parameters $k$, $t_1$, $t_2$, $\frac{M}{3}$ and $\mu$ are varied subject to the condition (\ref{ewbreakcondition}) and therefore we have effectively four free parameters.

\figh
At first,  we consider the right handed slepton mass $\tilde{m}_{\tilde{e}}$ which depends only on $k$, $t_1$ and $M$.
What we see later is that the experimental bound for $\tilde{m}_{\tilde{e}}$ gives a more stringent restriction on $t_1$ than other masses.
If we assume the off diagonal components of slepton mixing matrix are small,  the mass $\tilde{m}_{\tilde{e}}$ is
\begin{align}
\tilde{m}^2_{\tilde{e}}(\mu) \approx
\left[ -22 ( 1 +k )^2   \frac{{\alpha_1}^2(m_c)}{(4\pi)^2} +\left\{ 11 ( 1 +k )  -2 t_1 \right\}^2 \frac{2}{11}  \frac{\alpha_1^2(m_c) -\alpha_1^2(\mu)}{(4\pi)^2}\right] \left|\frac{M}{3}\right|^2 \nonumber\\
 -m^2_Z\cos 2\beta \sin^2\theta_W\label{RHSL}
\end{align}
Fig.\ref{figE}(a) shows the right handed slepton mass as the function of $k$ at $t_1=6$ and
Fig.\ref{figE}(b) shows the contour of $k$ and $t_1$ giving the mass the value of 94GeV, which is the present experimental lower bound.
In both Figs.\ref{figE}(a) and \ref{figE}(b), solid, dotted, broken and dashed lines correspond to $\frac{M}{3}=50$, 100, 500 and 1000TeV, respectively.
The allowed region in Fig.\ref{figE}(b) is above each line.

Secondly, we want to consider a constraint on $t_2$. 
It has turned out that the most stringent constraint on $t_2$ is obtained by the neutral scalar Higgs mass $\tilde{m}_{h^0}$.
Note that $\tilde{m}_{h^0}$ depends on $k$, $M$ and $t_2$ but hardly on $t_1$, 
and therefore we are able to get the constraint on $t_2$ irrespectively of the value $t_1$.
In Fig.\ref{figHA}(a) we show $\tilde{m}_{h^0}$ as the function of $k$ at $t_2=0$, 6 and 8, which correspond to solid, dotted and broken lines, fixing $\frac{M}{3}=50$TeV. 
Fig.\ref{figHA}(b) is the contour of $k$ and $t_2$ giving $\tilde{m}_{h^0}$ the  experimental lower bound 89.8GeV, where $\frac{M}{3}$ is varied as 50, 100, 500 and 1000TeV, and the allowed region is again above the lines.

\figMu

By looking at Figs.\ref{figE}(b) and \ref{figHA}(b), we are forced to choose the parameter $k$ around $-2$ to keep both $t_1$ and $t_2$ small simultaneously.
Fig.\ref{figE}(b) also show that the larger the parameter $\frac{M}{3}$ is, the smaller the parameters $t_1$ become.
However, if $\frac{M}{3}$ is grater than 500TeV we don't see much difference.
Moreover, the parameter $\frac{M}{3}$ is also constrained by the Higgs VEV relation (\ref{Vrelation}).
We have illustrated the parameter in Fig.\ref{figMu} as a function of $\mu$ by putting $v=245$GeV in (\ref{Vrelation}).
The solid and broken lines in Fig.\ref{figMu} correspond to $(k,t_1,t_2)=(-1.6,3,8)$, and $(k,t_1,t_2)=(-1.8,4,5)$, respectively.
If we suppose $\mu < 10$TeV, Fig.\ref{figMu} tells us to choose $\frac{M}{3}$ less than 500TeV.

\figt

Let us have a closer look at the regions $-2<k<-1.5$ in Figs.\ref{figE}(b) and \ref{figHA}(b), which are enlarged in Fig.\ref{figt}(a) and $t_1$ is drawn by bold lines.
We can easily see the condition to keep both $t_1$ and $t_2$ small simultaneously.
Fig.\ref{figt}(b)  shows the behavior of $t_1$ and $t_2$ as the function of $\frac{M}{3}$, where we have put $k=-1.6$, $-1.8$ and $-1.9$, corresponding to solid, dotted and broken lines.
In the case of $k=-2$ any region is allowed about $t_2$.
In agreement with our previous observation, both $t_1$ and $t_2$ are almost constant for a given $k$ and  $\frac{M}{3}>500$TeV.
For the purpose of surveying the allowed region of left handed slepton, Higgs, chargino and neutralino, we express $t_1$, $t_2$ and $\mu$ as the function of $k$ and $\frac{M}{3}$ with just experimental lower bound of the right handed slepton and neutral scalar Higgs mass.
We illustrate these masses in Figs.\ref{figL}-\ref{figChib}, using $t_1$, $t_2$ and $\mu$ as the function of $k$ fixing $\frac{M}{3}=50$, 100 and 500TeV and also as the function of $\frac{M}{3}$ fixing $k=-1.6$, $-1.8$ and $-1.9$, which correspond to solid, dotted and broken lines, respectively.
In all of the figures the allowed region is above each line and the dashed line denote the experimental lower bound.

In Fig.\ref{figL} we can see easily that $\tilde{m}_{\tilde{L}}$ exceeds the experimental bound in all the region of $k$ and $\frac{M}{3}$.
We show the mass spectra of $m_{H^0}$ in Fig.\ref{figH}.
In Fig.\ref{figH}(a) three lines degenerate and are constant.
This is explained as follows.
In Fig.\ref{figt} we vary $t_1$ and $t_2$ along each line so that $m_{h^0}$ is kept equal to 89.8GeV.
This indicates that ${m}^2_{A^0}=\bar{m}^2_{H_1}+\bar{m}^2_{H_2}$ is also constant since we know that $m_{h^0}$ and $m_{H^0}$ are dependent overtly on ${m}^2_{A^0}$.
This means variations of $t_1$ and $t_2$ do not affect ${m}^2_{A^0}$.
In other words to set $\tilde{m}_{h^0}=89.8$GeV is equivalent to fix the value of ${m}^2_{H^0}$ even if the parameters $t_1$, $t_2$, $k$ and $M$ are varied.
This is the reason why these lines degenerate and are constant.
In Fig.\ref{}(b) the two lines, corresponding to $k=-1.6$ and $-1.8$, also degenerate and are constant.
The reason is the same as above.
At $k=-2$ over $\frac{M}{3}=50$TeV, however, $t_2$ is allowed in any non-negative region.
We fix $t_2=0$ off the line in Fig.\ref{figt} so the behavior of broken lines gets different.
In Fig.\ref{figH} the experimental bound is exceeded.

\figL
\figH
Fig.\ref{figChia} shows chargino $m_{\tilde{\chi}^\pm_1}$ mass spectra and  
Fig.\ref{figChib} shows neutralino $m_{\tilde{\chi}^0_1}$ ones.
The other masses of the chargino and neutralino are heavier or have trivial mass spectrum. 
In these figures the lower bound of the allowed region have some V-shaped lines.
The reason is that we have plotted the absolute value of these masses, and these V-shaped lines go across the zero point.
Around the zero point the region of parameters are excluded by the experimental bound.
However, the other broad region is allowed.

After all we see that if a mass bounds of the right handed slepton and the neutral scalar Higgs are both considered, the other restriction from experimental bounds is almost satisfied.
This solves the tachyonic slepton problem.
Moreover, we have also discovered that for $-1.9<k<-1.6$ we are able to choose rather small values for $t_1$ and $t_2$.
This observation opens up a possibility to build a realistic model whose bilinear term's sector is not so much involved.
What we want to comment is that allowed region exists even if $t_2$ is equal to zero.
This means that we can build a model with only additional SU(2) singlet fields.
One of the simplest model is that there are three pairs of extra SU(2) singlet fields with a unit of U(1) charge which means $t_1=6$ and $t_2=0$, assuming $k\approx-2$ and $\frac{M}{3}>100$TeV. 

\figChia
\figChib

\clearpage

\section{Conclusion}\label{SecConclusion}
In the present paper, we have extended the anomaly mediation mechanism by including the effect due to K{\"a}hler anomaly.
We have thereby introduced a new parameter $k$, eq.(\ref{kdefinition}) which describes the effect of the K{\"a}hler potential.
We have also generalized the K{\"a}hler potential by including the bilinear terms whose effect is encoded in $t_l$ defined by eq.(\ref{tbilinear}), $l$ being the gauge group index.
We have derived new mass formulae (\ref{GauginomassFormula}) and (\ref{ScalarmassFormula}) which include the effects due to $k$ and $t_l$.

We have analyzed the allowed region of these parameters assuming the followings.
Firstly we assumed 50TeV $<\frac{M}{3}< 500$TeV.
This is because $\frac{M}{3}$ is related to the gravitino mass and therefore must be taken much larger than the present experimental  bound.
Secondly we set $k$ on the order of unity.
This sounds reasonable if the v.e.v. of $K_i F^i$ and $\frac{M}{3}$ come from the same origin.
Thirdly we assumed as small values as possible for $t_1$ and $t_2$.
This is simply for minimizing the number of extra fields $\phi_n$ with bilinear coupling.

We have analyzed all the masses listed in Table \ref{MassBound} and have found that the tachyonic slepton problem has been solved.
We have also discovered that for $-2<k<-1.5$ we are able to choose rather small numbers for $t_1$ and $t_2$.
This observation opens up a possibility to build a realistic model whose bilinear term's sector is not so much involved.
In particular we can set $t_2=0$ for $k \approx -2$.
This indicates a possibility to construct a model without SU(2) doublet field.
One of the simplest model is that there are three pairs of extra SU(2) singlet fields with a unit of U(1) charge. 

\section*{Acknowledgements}
I would like to thank Prof. Takahiro Kubota for stimulating discussions, many useful comments and collaboration.
I would also like to thank Dr. Tetsuo Shindou for physical and technical useful comments.

\end{document}